\documentclass[a4paper,11pt]{article}
\usepackage{pos}
\usepackage{lineno}

\ShortTitle{Study of collective phenomena via heavy quarks and quarkonia with ALICE}

\title{Study of collective phenomena via the production of heavy quarks and quarkonia in hadronic collisions}

\author*[a]{Victor Valencia Torres} 

\onbehalf{on behalf of ALICE collaboration}

\affiliation[a]{SUBATECH, IMT Atlantique, Nantes Université, CNRS/IN2P3\\
   Nantes, France}

\emailAdd{victor.valencia.torres@cern.ch}

\abstract{

Open heavy flavor and quarkonia have long been identified as ideal probes for understanding the quark-gluon plasma (QGP). Heavy quarks are produced in the early stage of the heavy-ion collisions. Therefore they experience the evolution of the medium produced, providing an important tool to investigate the properties of the QGP. In particular, the magnitude of the elliptic flow measured at the LHC is interpreted as a signature of the charm-quark thermalization in the QGP. This is  reflected in the azimuthal anisotropies of the final particles. In addition, the observation of collective-like effects in high-multiplicity pp and p--Pb collisions provides new insights on the evolution of QGP-related observables going from large to small collision systems. A better understanding of heavy-quark energy loss, quarkonium dissociation, and production mechanism can therefore be obtained with those system-size dependent observables. We present recent results of the $\mathrm{J}/\psi $ and open heavy-flavor hadrons flow in pp, p--Pb, and Pb--Pb collisions carried out by the ALICE collaboration.

}

\FullConference{%
  42st International Conference on High Energy physics - ICHEP2024\\
  18-24 July 2024 \\
  Prague, Czech Republic
}

\begin{document}

\maketitle

\section*{Introduction}
The production of heavy quarks, which occurs during initial hard-scattering processes of hadronic collisions, can be described using perturbative Quantum Chromodynamics (pQCD) calculations. In heavy-ion collisions, heavy-quark production is sensitive to the entire evolution of the QGP. Investigating the thermalization of heavy quarks within the QGP provides valuable  insights into its physical properties. By measuring open heavy-flavor hadrons and quarkonia, one can determine whether heavy quarks inherit flow from the QGP. Heavy quarks are also produced in smaller systems, where notable similarities have been observed at high $p_{\mathrm{T}}$ between high-multiplicity p--Pb and Pb--Pb collisions for observables typically associated with QGP signatures, such as long-range azimuthal correlations. These findings suggest the presence of collective phenomena in small systems ~\cite{AngularCorrelations}. Thus, investigating flow observables in the heavy-flavor sector is crucial for establishing and characterizing the collective behavior of charm and beauty quarks in small systems.  This can be done using azimuthal correlations of emitted particles, through the measurement of the elliptic flow ,\( v_2 \), in various colliding systems such as pp, p--Pb, and Pb--Pb.  In this contribution, we present the latest flow measurements of D mesons, of inclusive muons from hadron decays, and of quarkonia at different collision energies. We will also compare those results with theorical models.

The ALICE apparatus measures inclusive charmonia at forward rapidity ($2.5 < y < 4$) through the dimuon decay channel. At midrapidity, prompt and non-prompt heavy-flavor hadrons (with non-prompt originating from b-hadron decays) can be well distinguished. The key detectors at midrapidity include the Inner Tracking System (ITS), which is essential for tracking, vertexing, and measuring charged-particle multiplicity, and the Time Projection Chamber (TPC), which is utilized for tracking and particle identification by measuring specific energy loss. At forward rapidity, the muon spectrometer, consisting of a system of absorbers, a dipole magnet, and tracking and triggering stations, is used for reconstructing and identifying muon tracks. The V0 detector, which ensures a minimum-bias trigger, measures particle multiplicity, centrality and event-plane angle for the flow measurements. Additionally, inclusive hadrons at both forward and midrapidity can be measured down to zero $p_{\mathrm{T}}$. A detailed description of the ALICE detector is available in Ref. ~\cite{Alice}.

\section*{Results}

 The following section presents the $v_2$ measurements of heavy-flavor hadrons across various collision systems. These results provide new constraints on theoretical models in order to understand heavy-quark collectivity. Open heavy-flavor hadrons, composed of light and heavy quarks, are crucial for studying the transport coefficients of QGP. In the left panel of Fig. 1,  the average $v_2$ of prompt D\textsuperscript{0}, D\textsuperscript{+}, and D\textsuperscript{*+} mesons is compared  with the non-prompt D\textsuperscript{0}-meson $v_2$. The non-prompt D\textsuperscript{0}-meson $v_2$ is lower than the prompt non-strange D mesons with 3.2$\sigma$ significance in $2 < p_{\text{T}} < 8~\text{GeV}/c$, indicating  different degrees of thermalization between charm and beauty quarks in QGP. The weighted mean of the non-prompt D\textsuperscript{0}-meson $v_2$ in the measured $p_{\text{T}}$ range ($2 < p_{\text{T}} < 12~\text{GeV}/c$) is 2.7$\sigma$ above zero ~\cite{DMESONS}. In the right panel of Fig. 1,  the same non-prompt D\textsuperscript{0}-meson $v_2$ (blue) is compared with the $v_2$ of electrons from beauty-hadron decays (b$\rightarrow$ c $\rightarrow$ e)~\cite{bdecay}. Both measurements are also compared with theoretical models. The Linearized Boltzmann with diffusion model (LIDO) \cite{LIDO1} provides reasonable descriptions for these measurements, implementing beauty-quark transport in a hydrodynamically expanding QGP, including collisional and radiative processes. Beauty-quark hadronisation via coalescence is considered in addition to the fragmentation mechanism. The good agreement between the predictions for B-meson and non-prompt D\textsuperscript{0}-meson $v_2$ from LIDO indicates that the decay kinematics do not play a significant role in the beauty-hadron $v_2$ measurements. 

 \begin{figure}[h]
    \centering
    \includegraphics[width=0.85\linewidth]{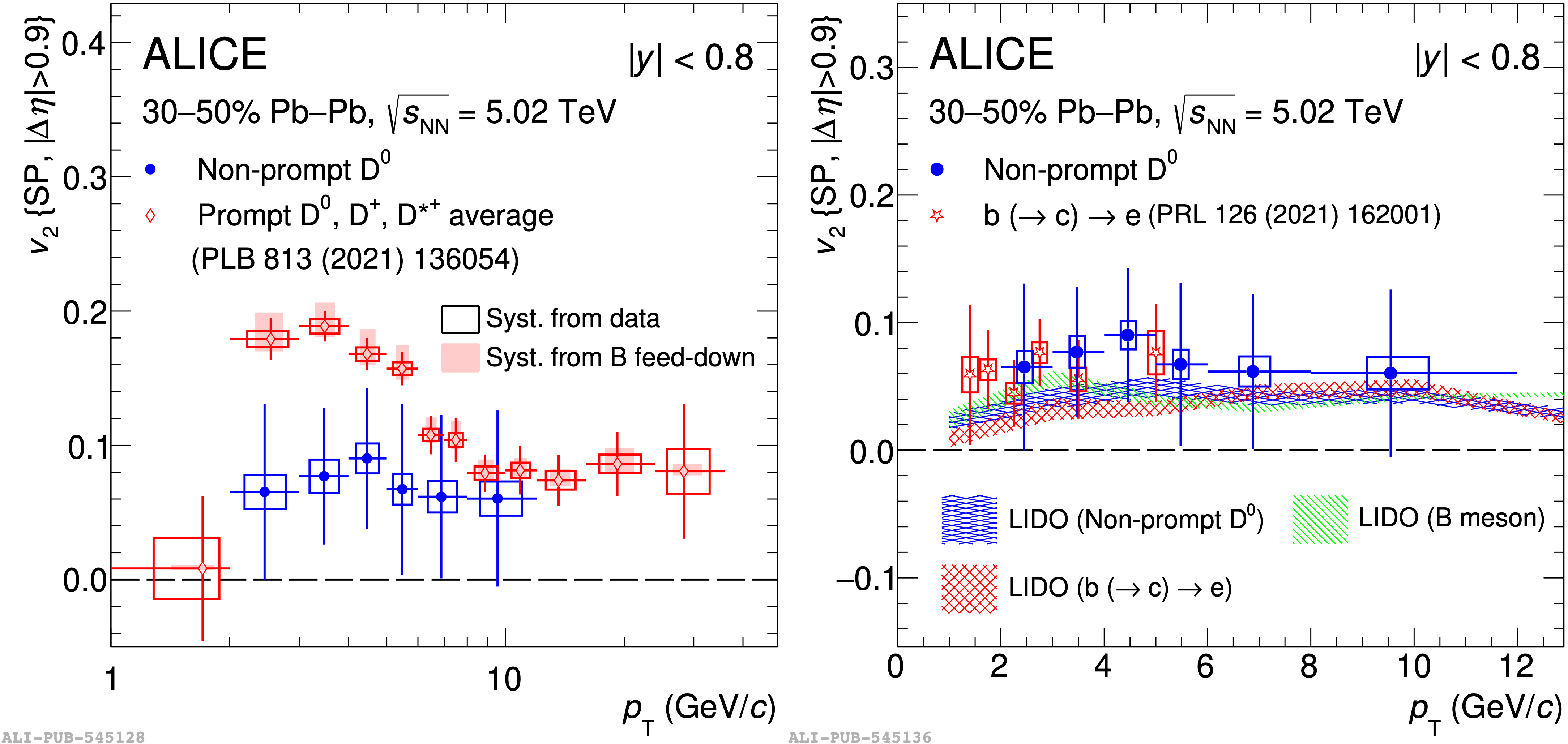}
    \caption{ Left: \( v_2 \) of non-prompt \( D^0 \) mesons (blue) and prompt non-strange \( D \) mesons (red) as a function of \( p_{\mathrm{T}} \) \cite{DMESONS}. Right: \( v_2 \) of non-prompt \( D^0 \) mesons \cite{DMESONS} (blue) and electrons from beauty-hadron decays \cite{bdecay} (red) as a function of \( p_{\mathrm{T}} \). Both measurements are measured in the 30--50\%  centrality class in Pb--Pb collisions and are compared with LIDO model predictions \cite{LIDO1}, \cite{LIDO2}.}
    \label{fig:example}
\end{figure}

 These measurements will provide important constraints to model predictions, allowing to extract accurately the spatial diffusion coefficient of beauty quark in QGP. Another way to study experimentally the flow in the heavy-quark sector is by using inclusive muons from heavy-flavor hadron decays. Figure 2 presents the $p_{\mathrm{T}}$-differential elliptic flow coefficient, $v^{\mu}_{2}$, of inclusive muons, with non-flow effects subtracted in the 0--20\% multiplicity class for p--Pb collisions at $\sqrt{s_{\mathrm{NN}}} = 8.16$ TeV.

\begin{figure}[h]
    \centering
    \includegraphics[width=1.0\linewidth]{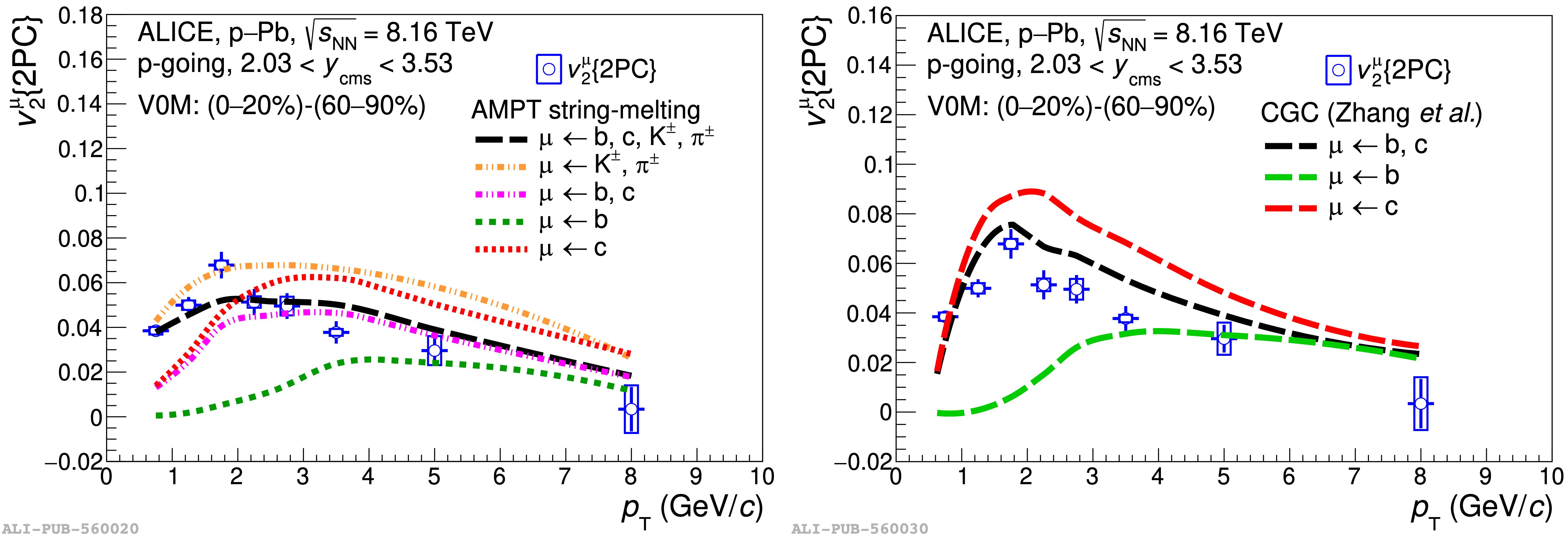}
    \caption{Flow of inclusive muon $v_2^{\mu}$ as a function of $p_{\mathrm{T}}$ at forward rapidity in high-multiplicity p--Pb collisions at $\sqrt{s_{\mathrm{NN}}} = 8.16$ TeV.}
    \label{fig:example}
\end{figure}
This new measurement covers for the first time a wide transverse momentum range of $0.5 < p_{\mathrm{T}} < 10$ GeV/$c$ in the forward rapidity region. The data reveal an initial increase in $v^{\mu}_{2}$ with rising $p_{\mathrm{T}}$, reaching a peak value of approximately 0.07 at $p_{\mathrm{T}} \sim 2$ GeV/$c$, followed by a decrease as $p_{\mathrm{T}}$ continues to increase. The region of particular interest for studying the flow in the heavy-quark sector is $p_{\mathrm{T}} > 2$ GeV/$c$, where muons predominantly originate from heavy-flavor hadron decays and surpass contributions from light-hadron decays (such as charged-pion and -kaon decays at low $p_{\mathrm{T}}$).
 The results are compared to a multi-phase transport (AMPT) \cite{AMPTmodel} and color-glass condensate (CGC) \cite{CGCmodel} models in the left and right panel, respectively. Both models describe the data within uncertainties. The measurements were performed using two-particle correlations $v^\mu_{2}\{2\mathrm{PC}\}$ \cite{InclusiveMuons}. Statistical uncertainties (represented by vertical bars) were estimated using the Fourier fit method, while systematic uncertainties are depicted by empty boxes. These results suggest that heavy quarks exhibit significant flow at mid-to-high $p_{\mathrm{T}}$ in p--Pb collisions.

 In order to study the contribution  of charmonia in collective effects in hadronic collisions, the elliptic flow $v_{2}$ of J/$\psi$ was measured across various colliding systems.

\begin{figure}[h]
    \centering
    \includegraphics[width=0.69\linewidth]{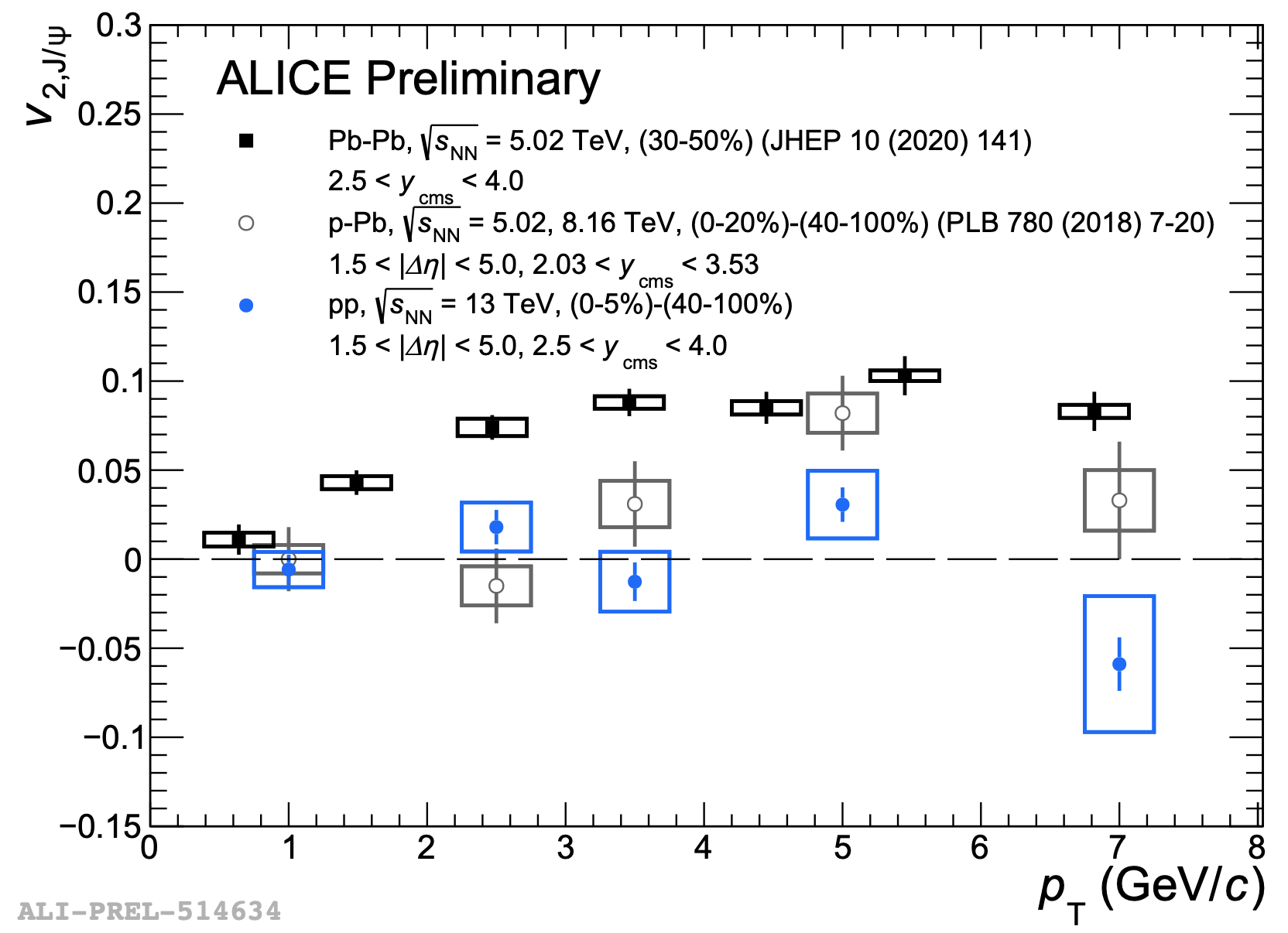}
    \caption{The \( p_{\mathrm{T}} \) dependence of $v_2^{\mathrm{J}/\psi}$ in pp, p--Pb \cite{p--Pbjpsi}, and Pb--Pb \cite{Pb--Pbjpsi} collision systems.}
    \label{fig:example}
\end{figure}

In Fig. 3, the flow measurement in pp is presented and compared with p--Pb \cite{p--Pbjpsi} and Pb--Pb \cite{Pb--Pbjpsi} flow data. The $v_2$ measurement in high-multiplicity pp and p--Pb collisions is obtained using azimuthal correlations between inclusive dimuons and charged hadrons. The Pb--Pb measurement is obtained using the scalar product method (SP). For p--Pb collisions, the elliptic flow of J/$\psi$  exhibits similar values to those observed in Pb--Pb collisions when $p_{\mathrm{T}} > 4$ GeV/$c$, suggesting a shared mechanism for generating flow in both collision systems. Nonetheless, the TAMU transport model \cite{RegenerationCharmonia}, which successfully describes $v_2$ in Pb--Pb collisions, interprets this observation as a result of energy loss. At low $p_{\mathrm{T}}$, the flow contribution is understood to result from the regeneration of J/$\psi$. Alternatively, other theoretical frameworks based on initial-state models, like CGC, predict that correlations among final-state particles could produce flow-like effects \cite{collectivitySmall}. In pp collisions, the J/$\psi$ $v_2$ does not significantly deviate from zero as a function of $p_{\mathrm{T}}$. This finding holds true even when considering the $v_2$ integrated over $p_{\mathrm{T}}$ in the range $1 \leq p_{\mathrm{T}} \leq 12$ GeV/$c$, where the deviation from zero is within $1\sigma$. Therefore, within the current uncertainties, no collective behavior is observed for the J/$\psi$ in high-multiplicity pp collisions at the LHC. Finally, the significantly larger Run 3 dataset as compared to Run 1 and 2 will enable more precise flow measurements, providing unprecedented breakthrough for the understanding of the flow of heavy quarks in hadronic collisions.

\begin{figure}[h]
    \centering
    \includegraphics[width=0.99\linewidth]{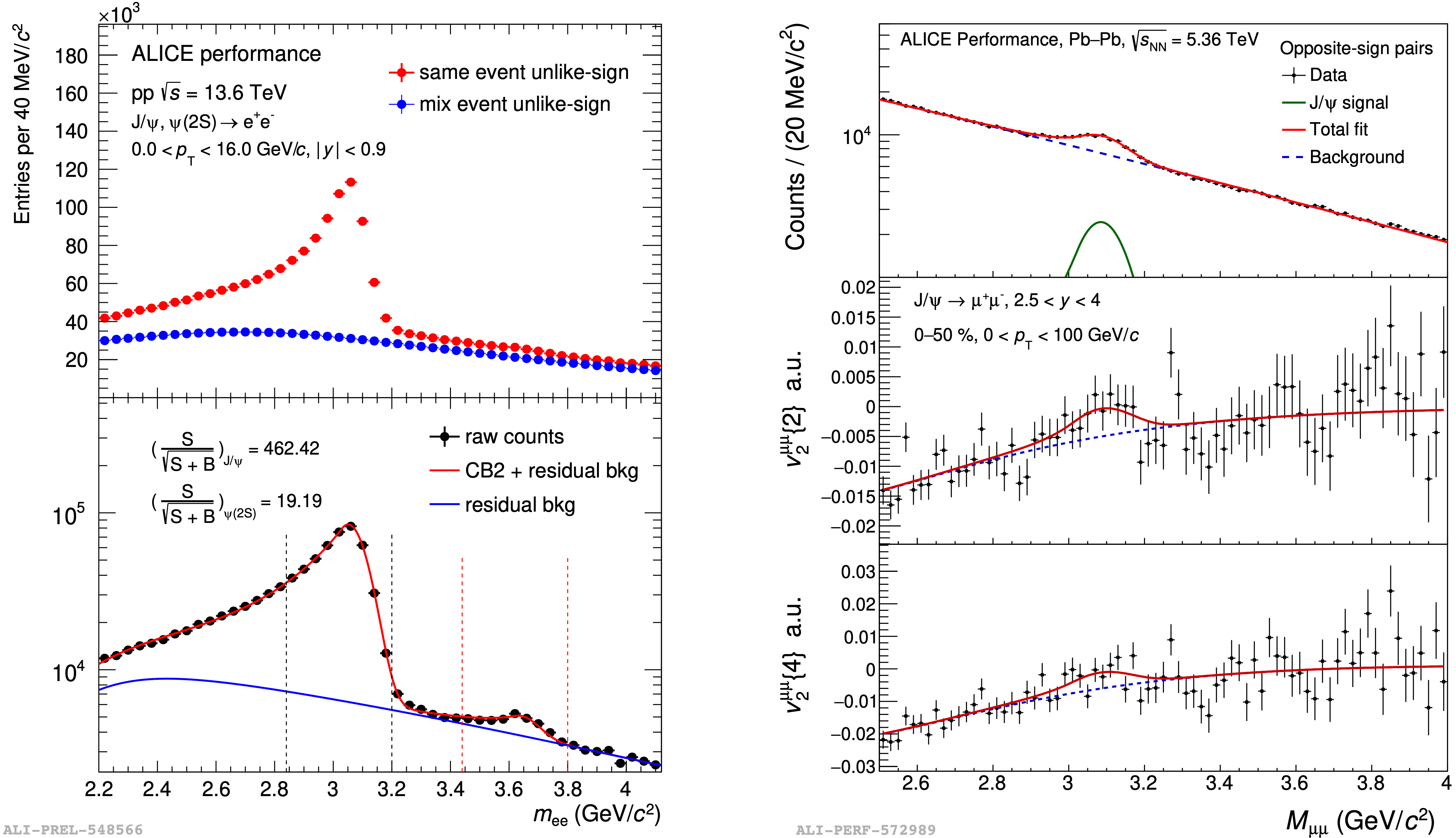}
    \caption{Left: Fit of the invariant mass distribution $m_{ee}$ at midrapidity in pp collisions at \( \sqrt{s}= 13.6 \, \text{TeV} \),  showing the extracted inclusive $\mathrm{J}/\psi$ and $\psi(\mathrm{2S})$ signals.
    Right: Fits of $m_{\mu\mu}$, $v_2^{\mu\mu}\{2\}$ and $v_2^{\mu\mu}\{4\}$ distributions  at forward-rapidity in Pb--Pb collisions at \( \sqrt{s_{\mathrm{NN}}}= 5.36 \, \text{TeV} \), where the inclusive $\mathrm{J}/\psi$, $v_2^{\mathrm{J}/\psi}\{2\}$ and $v_2^{\mathrm{J}/\psi}\{4\}$ are also extracted (fraction of the total Pb--Pb statistics).   
    }
    \label{fig:run3data}
\end{figure}

The left panels of Fig. 4 show the signal extraction of inclusive $\mathrm{J}/\psi$ and  $\psi(\mathrm{2S})$ from the $m_{ee}$ distribution at midrapidity ($|y| < 0.9$) in pp collisions at \( \sqrt{s}= 13.6 \, \text{TeV} \), with tracks selected by $p_{\mathrm{T}} > 1$ GeV/c. An event-mixing technique was applied for unlike-sign di-electrons. In the right panels of Fig. 4, the signal extraction of inclusive $\mathrm{J}/\psi$ from the $m_{\mu\mu}$ distribution at forward rapidity ($ 2.5  <y < 4$) in Pb--Pb collisions at \( \sqrt{s_{\mathrm{NN}}}= 5.36 \, \text{TeV} \) is presented on top. Additionally, the two bottom right panels show the multi-particle cumulant coefficients $v_2^{\mu\mu}\{2\}$ and $v_2^{\mu\mu}\{4\}$ as function of $m_{\mu\mu}$.

\section*{Summary}

 We have shown that non-prompt \(D^0\) mesons exhibit a lower \(v_2\) compared to prompt \(D\) mesons in Pb--Pb collisions, indicating different degrees of thermalization between charm and beauty quarks within the QGP. In p--Pb collisions, a notable elliptic flow is observed for inclusive muons at $p_{\mathrm{T}} > 2$ GeV/$c$, highlighting the contribution of heavy quarks to collective effects. For \(\mathrm{J}/\psi\) mesons, the elliptic flow in p--Pb collisions shows similarities to Pb--Pb collisions, suggesting a shared mechanism for generating flow. However, in high-multiplicity pp collisions, the \(\mathrm{J}/\psi\) \(v_2\) remains close to zero, indicating the absence of significant collective behavior in these smaller collision systems. Finally, using Run 3 data, we present the first signal extraction of \(\mathrm{J}/\psi\) at mid and forward rapidity in pp and Pb--Pb collisions, respectively. At forward-rapidity, a new "multiparticle cumulant" method is shown for the first time to extract $v_2^{\mathrm{J}/\psi}$ with improved precision, reducing non-flow effects.

\end{document}